\documentclass[prl,aps,twocolumn,showpacs,preprintnumbers,amsmath,amssymb]{revtex4}

\usepackage{graphicx}
\usepackage{bm}
\usepackage{epsfig}
\usepackage{color}

\begin{document}

\preprint{}

\title{Evidence for multi-band strongly coupled superconductivity  in SmFeAsO$_{0.8}$F$_{0.2}$ single crystals by high-field vortex torque magnetometry}

\author{L.\ Balicas,$^1$ A.\ Gurevich,$^1$ Y.\ J.\ Jo,$^1$ J.\ Jaroszynski,$^1$ D.\ C.\ Larbalestier,$^1$
R. H. Liu,$^2$ H. Chen,$^2$ X. H. Chen,$^2$ N.\ D.\ Zhigadlo,$^3$
S.\ Katrych,$^3$ Z.\ Bukowski,$^3$ and J.\ Karpinski,$^3$}

\affiliation{$^1$National High Magnetic Field Laboratory, Florida
State University, Tallahassee-FL 32310, USA}
\affiliation{$^2$Hefei National Laboratory for Physical Science a
Microscale and Department of Physics, University of Science and
Technology of China, Hefei, Anhui 230026, People's Republic of
China} \affiliation{$^3$Laboratory for Solid State Physics, ETH
Z\"{u}rich, CH-8093 Z\"{u}rich, Switzerland}

\date{\today}%
\begin{abstract}
To probe manifestations of multiband superconductivity in
oxypnictides, we measured the angular dependence of magnetic
torque $\tau(\theta)$ in the mixed state of
SmO$_{0.8}$F$_{0.2}$FeAs single crystals as functions of
temperature $T$ and high magnetic field $H$ up to 30 T. We show
that the effective mass anisotropy parameter $\gamma$ extracted
from $\tau(\theta)$, can be greatly overestimated if the strong
paramagnetism of Sm or Fe ions is not properly taken into account.
The correctly extracted $\gamma$ depends on both $T$ and $H$,
saturating at $\gamma \simeq 9$ at lower temperatures. Neither the
London penetration depth nor the superfluid density is affected by
high fields fields up to the upper critical field. Our results
indicate two strongly-coupled superconducting gaps of nearly equal
magnitudes.
\end{abstract}

 \maketitle

The recently discovered superconducting oxypnictides
\cite{discovery,chen} have similarities with the high $T_c$
cuprates, such as the emergence of superconductivity upon doping a
parent antiferromagnetic compound \cite{chen,phasediagram,AFM}.
Several theoretical models \cite{theory1, theory2} suggest
unconventional superconducting pairing, while the Andreev
spectroscopy \cite{pcs}, penetration depth
\cite{penetrationdepth}, and photoemission measurements
\cite{ding} indicate nodless s-wave pairing symmetry. Experiments
\cite{ding, hunte, weyeneth} have found evidence for multi-gap
superconductivity, in agreement with theoretical predictions
\cite{theory1}.

The comparatively high $T_c$ values  and extremely high upper
critical fields $H_{c2}$ of the oxypnictides \cite{hunte,jo}
indicate promising prospects for technological applications if,
unlike the layered cuprates, a sizeable vortex liquid region
responsible for dissipative flux flow does not dominate their
temperature-magnetic field $(T-H)$ phase diagram. It is therefore
important to reveal the true behavior of the anisotropic
magnetization in the vortex state of the oxypnictides, particularly
the extent to which  vortex properties are affected by strong
magnetic correlations, multiband effects and possible interband
phase shift between the order parameters on different pieces of the
Fermi surface \cite{theory1}. For instance, multiband effects in
MgB$_2$ can manifest themselves in strong temperature and field
dependencies for the mass anisotropy parameter $\gamma(T,H)$ and the
London penetration depth $\lambda(T,H)$ even at $H\ll H_{c2}$
\cite{MgB2vsT, MgB2vsH}. Yet, there are significant differences
between two-band superconductivity in MgB$_2$ and in oxypnictides:
in MgB$_2$ the interband coupling is weak, while in the oxypnictides
it is the strong interband coupling which is expected to result in
the high $T_c$ \cite{theory1}. Thus, probing multiband
superconductivity in oxypnictides by magnetization measurements
requires high magnetic fields, which can suppress the superfluid
density in the band with the largest coherence length above the
``virtual upper critical field" ($H_v$) at which the vortex cores in
this band overlap. In this Letter we address these issues,
presenting the first high-field torque measurements of anisotropic
reversible magnetization of the vortex lattice in
SmO$_{0.8}$F$_{0.2}$FeAs single crystals. Our measurements of
$\gamma(T,H)$ up to 30T and extended temperature range, $20<T<40$ K
have revealed a different behavior of $\gamma(T,H)$ as compared to
recent low-field torque measurements \cite{weyeneth}.

Measurements of anisotropic equilibrium magnetization $m(T,H)$ in
SmO$_{0.8}$F$_{0.2}$FeAs are complicated by the smallness of
$m(H,T)$ caused by the large Ginzburg-Landau parameter,
$\kappa=\lambda/\xi > 100$ and by the strong paramagnetism of
Sm$^{3+}$ ions, which can mask the true behavior of $m(T,H)$.  In
this situation torque magnetometry is the most sensitive technique
to measure the fundamental anisotropy parameters of $\bf{m}(T,H)$
in small single crystals. The torque $\tau={\bf m}\times{\bf H}$
acting upon a uniaxial superconductor is given by
    \begin{equation}
    \tau(\theta) = \frac{HV \phi_0 (\gamma^2-1)\sin2\theta}{16\pi \mu_0 \lambda_{ab}^2\gamma\varepsilon(\theta)}
    \ln \left[ \frac{\eta H_{c2}^{ab}}{\varepsilon(\theta) H}\right]+\tau_m\sin 2\theta,
    \label{Kogan}
    \end{equation}
where $V$ is the sample volume, $\phi_0$ is the flux quantum,
$H_{c2}^{ab}$ is the upper critical field along the ab planes,
$\eta\sim 1$ accounts for the structure of the vortex core,
$\theta$ is the angle between $\bf{H}$ and the c-axis,
$\varepsilon(\theta) = (\sin^2 \theta+\gamma^2\cos^2\theta)^{1/2}$
and $ \gamma = \lambda_c/ \lambda_{ab}$ is the ratio of the London
penetration depths along the c-axis and the ab-plane. The first
term in Eq. (\ref{Kogan}) was derived by Kogan in the London
approximation valid at $H_{c1}\ll H\ll H_{c2}$ \cite{kogan}. The
last term in Eq. (1) describes the torque due to paramagnetism of
the SmO layers and possible intrinsic magnetism of the FeAs
layers. Here $\tau_m=(\chi_c-\chi_a)VH^2/2$ and $\chi_c$ and
$\chi_a$ are the normal state magnetic susceptibilities of a
uniaxial crystal along the c-axis and ab plane, respectively.  As
will be shown below, the paramagnetic term in Eq. (\ref{Kogan}) in
SmO$_{0.8}$F$_{0.2}$FeAs can be larger than the superconducting
torque, which makes extraction of the  equilibrium vortex
magnetization rather nontrivial. In this Letter we develop a
method, which enables us to resolve this problem and measure the
true angular dependence of the superconducting torque as a
function of both field and temperature, probing the concomitant
behavior of $\gamma(T,H)$ and $\lambda_{ab}(T,H)$ and
manifestations of multiband effects in SmO$_{0.8}$F$_{0.2}$FeAs
single crystals.
\begin{figure}[htb]
\begin{center}
\epsfig{file= 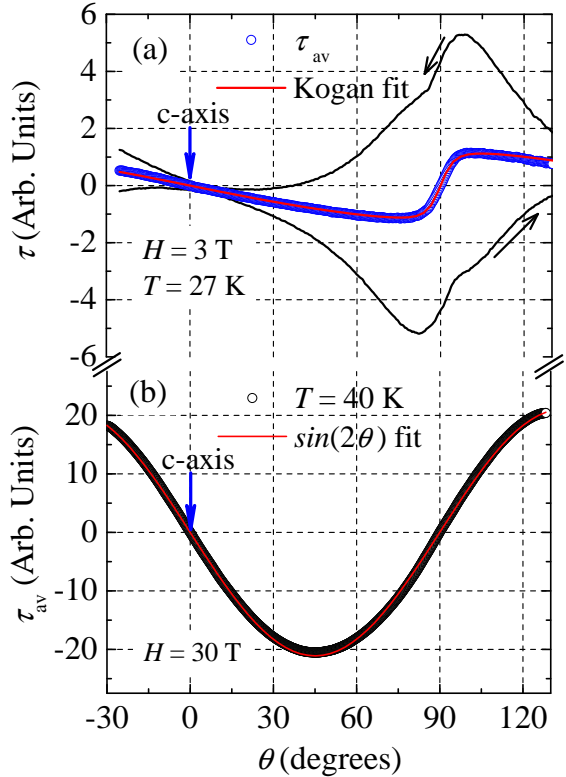, width = 7.4 cm} \caption {(color
online) (a) Magnetic torque $\tau(\theta)$ for a
SmO$_{0.8}$F$_{0.2}$FeAs single crystal for increasing and
decreasing angle sweeps (black lines) at 3 T and 27 K. The
equilibrium $\tau_{av}(\theta)$ (blue markers) is obtained by
averaging both traces. (b) $\tau_{av}(\theta)$ for 40 K and 30T
exhibits a nearly sinusoidal angular dependence. Red line
corresponds to a fit to the first term in Eq. (1) with $\gamma =
11.5$.}
\end{center}
\end{figure}

Underdoped single crystals of SmO$_{1-x}$F$_x$FeAs having typical
sizes, $100 \times 100\times 10 $ $\mu$m$^3$ and $T_c \simeq 45$K
were grown by the flux method described in Ref. \cite{zhigadlo}.
The sample was attached to the tip of a piezo-resistive
micro-cantilever placed into a rotator inserted into a vacuum can.
The ensemble was placed into a $^4$He cryostat coupled to a
resistive 35 T dc magnet of the National High Magnetic Field Lab.
Changes in the resistance of the micro-cantilever associated with
its deflection and thus a finite magnetic torque $\tau$ was
measured via a Wheatstone resistance bridge.

\begin{figure}[htb]
\begin{center}
\epsfig{file= 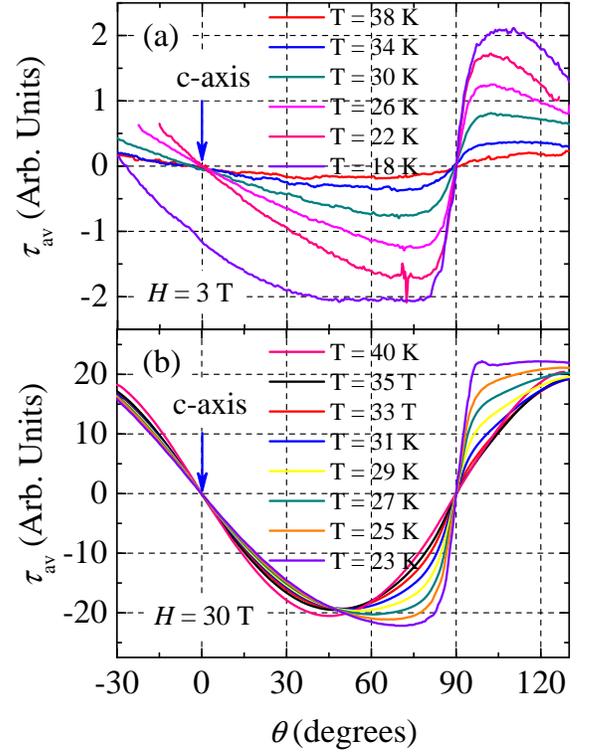, width = 7.4 cm} \caption {(color
online) (a) Angular dependence of $\tau_{\text{av}}(\theta)$ at 3
T and several temperatures. (b) Same as in (a) for $30$ T. }
\end{center}
\end{figure}

Fig. 1 (a) shows typical angular dependence of $\tau(\theta)$ at 27
K and 3 T. Since we are only interested in temperature and field
dependencies of $\gamma$ and $\lambda$, the torque data are provided
in arbitrary units. A hysteresis, resulting from the irreversible
magnetization is observed between increasing and decreasing angle
sweeps. Black markers depict the average value of both traces,
$\tau_{\text{av}}(\theta) = (\tau_\uparrow(\theta)+
\tau_\downarrow(\theta))/2$ defined as an equilibrium magnetization,
where the arrows indicate increasing or decreasing angle sweeps. The
red line is a fit to the first term in Eq. (1) with $\gamma \approx
11.5$ a value that is $\simeq 25$\% smaller than the value reported
in Ref. \cite{weyeneth}. However, this multiparameter fit is not
very suitable for extraction of the true values of $\gamma$ due to
pronounced error bars for $H_{c2}^{ab}$ and a significant
paramagnetic component particularly at 30T. The complete set of the
raw $\tau_{av}(\theta)$ data is shown in Figs. 2 (a) and (b).

The superconducting component of the torque can be unambiguously
extracted from the data by fitting the sum of two measured curves
$\tau_{\text{av}}(\theta)+\tau_{\text{av}}(\theta + 90^{\circ})$,
in which the paramagnetic component cancels out:
    \begin{eqnarray}\label{Koga}
    \tau(\theta)+ \tau(\theta + 90^{\circ}) = \frac{V \phi_0(\gamma^2-1) H\sin2 \theta}{16\pi \mu_0 \lambda_{ab}^2\gamma}
    \nonumber \\
    \times \left[ \frac{1}{\varepsilon(\theta)} \ln \left(
    \frac{\eta H_{c2}^{ab}}{\varepsilon(\theta) H}
    \right) - \frac{1}{\varepsilon^{\star}(\theta)} \ln \left( \frac{\eta H_{c2}^{ab}}{\varepsilon^{\star}(\theta) H}
    \right) \right],
    \end{eqnarray}
where $\varepsilon^{\star}(\theta) = ( \cos^2\theta + \gamma^2
\sin^2\theta )^{1/2}$.
\begin{figure}
\begin{center}
\epsfig{file=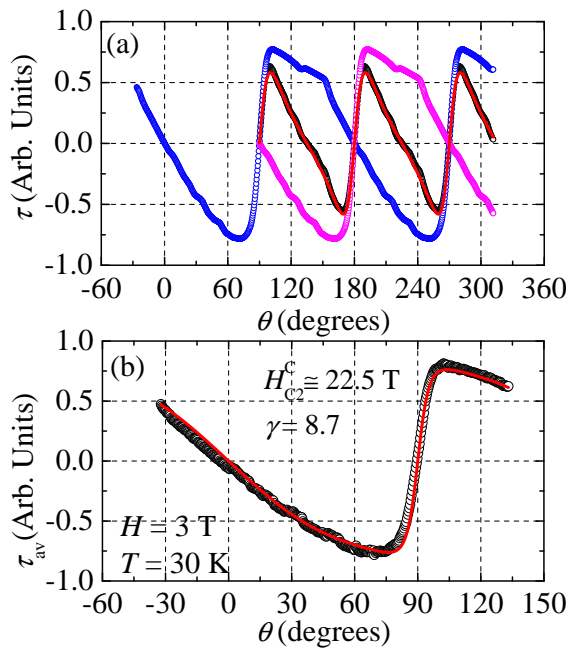, width=7.4 cm} \caption{(color
online) Angular dependence of $\tau_{\text{av}}(\theta)$ (blue)
and $\tau_{\text{av}}(\theta + 90^{\circ})$ (magenta) for 3 T and
30 K. Red line corresponds to a fit of
$\tau_{\text{av}}(\theta)+\tau_{\text{av}}(\theta + 90^{\circ})$
to Eq. (2). (b) An example of the fit of
$\tau_{\text{av}}(\theta)$ to Eq. (1) with the parameters taken
from Fig. 3 (a).}
\end{center}
\end{figure}
\begin{figure}
\begin{center}
\epsfig{file=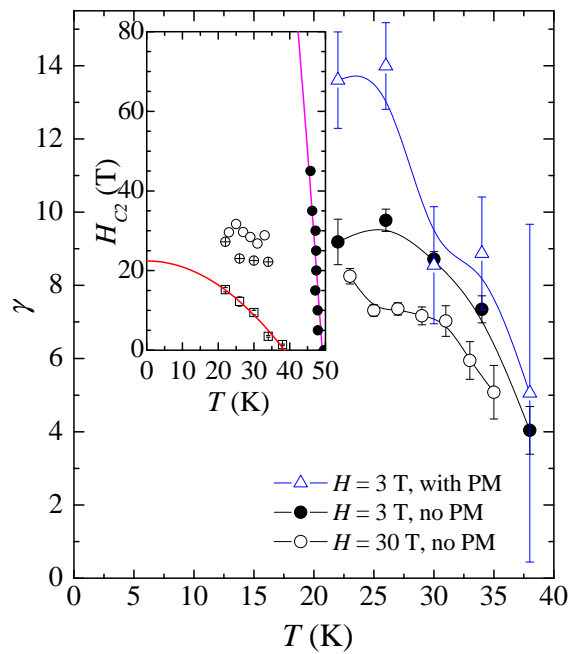, width=7.4 cm} \caption{(color
online) Temperature dependencies of $\gamma$ obtained from fitting
$\tau_{av}(\theta)+ \tau_{av}(\theta+90^{\circ})$ to Eq. (2) for
3T (solid circles) and 30T (open circles). The ambiguity of the
extraction of $\gamma$ from a direct multiparameter fit of
$\tau_{av}(\theta)$ \cite{weyeneth} to the first term in Eq. (1)
is illustrated by the triangles. Inset shows the difference of
$H_{c2}$ for different fit procedures. Squares correspond to $\eta
H_{c2}^{c}$ obtained by fitting the raw $\tau_{av}(\theta, T, H =
3 \text{T})$ solely to first term in Eq. (1). Solid circles
correspond to $H_{c2}^{ab}$ measured on a polycrystalline
SmO$_{0.8}$F$_{0.2}$FeAs with a $T_c \sim 47$ K \cite{jo} which we
used to fit the data to Eq. (2) and from which we obtain
$H_{c2}^c=H_{c2}^{ab}/\gamma$ (open circles are points from $H = 3$ T while crossed circles from $H = 30$ T).}
\end{center}
\end{figure}
\begin{figure}[htp]
\begin{center}
\epsfig{file=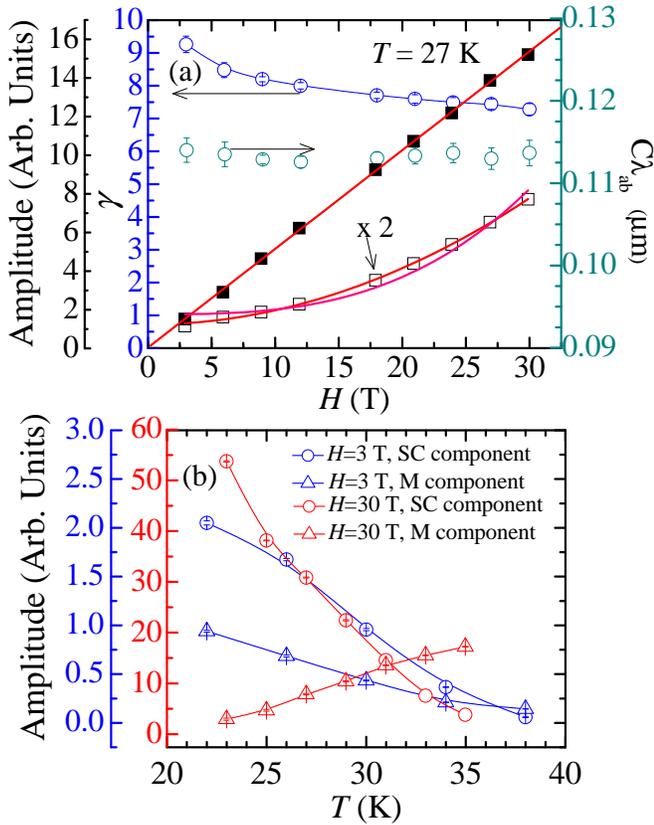, width=8.6 cm} \caption{ (color
online) (a) Field dependence of $\gamma$ (blue markers) and $C \lambda_{ab}$ (green markers)
obtained by fitting ($\tau(\theta)+ \tau(\theta+90^{\circ})$ at
$T = 27$ K to Eq. (2). Here, $C$ is a calibration dependent constant for our cantilever. Black markers depict the field dependence
of respectively the magnetic (open squares) and superconducting (solid squares) components. For clarity the data was multiplied by a factor of two.
Red line corresponds to a $T^2$ fit and the magenta line to a $T^3$ dependence.
(b) Amplitudes of superconducting
(circles) and magnetic (triangles) components of
$\tau_{av}(\theta)$, at $H = 3$ T (blue markers) and 30 T (red
markers). }
\end{center}
\end{figure}
This procedure is illustrated by Fig. 3 (a) where $\tau_{\text{av}}(
\theta)$ for 3T and 30K is plotted together with
$\tau_{\text{av}}(\theta+90^{\circ})$. Black markers depict the sum
of both traces which is entirely determined  by the superconducting
response. Red line corresponds to a fit to Eq. (\ref{Koga}), where
$\eta H_{c2}^{ab}[T]\approx 315(1-T^2/T_c^2) $ shown in the inset of
Fig. 4 was extracted from the onset of the resistive transition in
polycrystalline SmO$_{0.8}$F$_{0.2}$FeAs with $T_c \simeq 47 K$
\cite{jo}. The onset of the resistive transition reflects the
behavior of those crystallites having the field along the ab-plane
thus $H_{c2}^{ab}(T)$. Given the lack of high-field $H_{c2}(T)$ data
for our single crystals, the use of the measured value $H_{c2}^{ab}$
for polycrystals eliminates the ambiguities of the multiparameter
fit if $H_{c2}$ in Eq. (2) is treated as one more fit parameter.
Using the superconducting parameters obtained by this method in Eq.
(1) we then fit the original $\tau_{\text{av}}(\theta)$ (red line in
Fig. 3 (b)), treating $\tau_m$ as the only adjustable parameter.
This method does result in an excellent fit, allowing us to extract
both the superconducting torque and the paramagnetic torque
associated with Sm or Fe moments. It also gives smaller values for
$\gamma(T)$ than the ones obtained by direct fitting
$\tau_{\text{av}}(\theta)$ uniquely to the first term of Eq.
(\ref{Kogan}). As shown in Figs. 4 and 5 (a), the so-obtained
$\gamma$ is not only temperature dependent \cite{weyeneth} but it is
also field dependent, decreasing by more than 20 \% at 30 T.

The obtained temperature dependence of $\gamma(T,H)$ is reminiscent
of the behavior of $\gamma(T,H)$ previously reported for MgB$_2$
which was explained in terms of multi-band effects \cite{MgB2vsT,
MgB2vsH}. Yet the extracted London penetration depth shown in Fig. 5
(a) does not exhibit a significant field dependence which would
indicate an abrupt depression of the superfluid density in one of
the bands above $H_v$ (produced by the suppression of the respective
superconducting gap). This is quite remarkable given that we
measured $\lambda (T,H)$ up to the applicability limit of the London
theory, i.e. $\eta H_{c2}^c/H \simeq 1$ at $H=30$ T. Overall, the
behaviors of $\gamma(T,H)$ and $\lambda(T,H)$ shown in Fig. 5 (a)
would suggest two strongly coupled gaps of similar magnitude but not
too different mass anisotropies. The decrease of $\gamma(H)$ as $H$
increases may indicate that the band with the shorter coherence
length $\xi\sim \hbar v_F/\Delta$ is the least anisotropic.

The relative contributions of the superconducting and magnetic
components in $\tau_{\text{av}}(\theta, T,H)$ are shown in Fig. 5
(b). At higher $T$ the behavior of $\tau_m(T) \propto C_1/T[K]+C_2
$, $C_2\approx -C_1/43$ at 3T is consistent with the Curie-Weiss
paramagnetism of Sm$^{3+}$ ions. However, this temperature
dependence of $\tau_m(T)$ changes at $H = 30$ T, for which
$\tau_m(T)$ decreases as $T$ decreases, indicating the intriguing
possibility of field-induced antiferromagnetism (AF). Indeed, the
coexistence of AF and superconductivity in phase-separated regions
has been reported in the oxypnictides \cite{goko}, while AF in the
vortex cores of the cuprates has been observed at higher field
\cite{cuprates}. In our case, the latter can be ruled out since the
observed $\tau_m(H,T=27 \text{K})$ follows a $H^2$ dependence,
unlike $\tau_m\propto H^3/H_{c2}$ for the AF cores. Another
mechanism may result from a field-dependent uniaxial magnetic
anisotropy as the Sm$^{3+}$ moments align along the field. This
effect can be modeled by the single-ion Hamiltonian
$\hat{H}=\mu_B(g_a\sigma_xH_x+g_a\sigma_yH_y+g_c\sigma_zH_z)/2$,
where  $\sigma_\alpha$ are the Pauli matrices, and $g_\alpha$ are
the principal values of the effective $g$-factor tensor \cite{mag}.
In this case
    \begin{equation}
    \tau_p=n\mu_B(g_c^2-g_a^2)H\sin 2\theta\tanh(\mu_Bg_\theta H/2T)/2g_\theta,
    \label{anis}
    \end{equation}
where $n$ is the density of paramagnetic ions,
$g_\theta=(g_c^2\cos^2\theta+g_a^2\sin^2\theta)^{1/2}$. For weak
fields or $\mu_Bg_\theta H\ll T$, Eq. (\ref{anis}) gives
$\tau_p=(\chi_c-\chi_a)H^2\sin 2\theta /2$ and
$\chi_\alpha=\mu_B^2g_\alpha^2/4T$ as used in our analysis.
However, for higher fields $\mu_Bg_\theta H> T$, the paramagnetic
torque $\tau_p\simeq n\mu_B(g_c^2-g_a^2)H\sin 2\theta/2g_\theta$
acquires higher order harmonics. This case may pertain to our data
at 30T and $T<30$K, for which the pure $\sin 2\theta$ component in
$\tau(\theta)$ does decrease as $T$ decreases, but
$\tau_p(\theta)$ may not be completely eliminated by the procedure
described above. Deviations from Eq. (\ref{Kogan}) also come from
corrections to the London theory at high fields resulting in
additional terms $\propto \alpha H_{c2}/H-(\ln\eta
+\alpha)H/H_{c2}, \ \alpha \sim1$ in ${\bf m}(T,H,\theta)$ due to
pairbreaking and nonlocal effects \cite{core}.

In summary, our torque measurements at high-fields reveal the
temperature and field dependencies of the anisotropic reversible
magnetization which is strongly coupled with the magnetism of rare
earth ions in SmO$_{0.8}$F$_{0.2}$FeAs single crystals. Our results
indicate a temperature and field dependent mass anisotropy
$\gamma(T,H)$ which saturates at $\gamma \simeq 9$ at low
temperatures under a modest field. This value is higher than $\gamma
= H_{c2}^{c}/ H_{c2}^{ab} \simeq 5$ at low temperatures in
NdO$_{0.7}$F$_{0.3}$FeAs single crystals \cite{jan} and $\gamma
\simeq 5-7$ for YBa$_2$Cu$_3$O$_{7-\delta}$, but is much smaller
than $\gamma \geq 30$ suggested by Ref. \cite{dubroka}. The observed
insensitivity of the London penetration depth at fields up to 30T is
indicative of strong coupling superconductivity, which in addition
to a not very high $\gamma$ is very important for applications. Our
results are consistent with strongly coupled gaps of nearly equal
magnitudes in distinct bands.

The NHMFL is supported by NSF through NSF-DMR-0084173 and the
State of Florida. This work was also supported by NHMFL/IHRP (LB,
AG, DCL), NHMFL-Schuller program (YJJ) and by AFOSR (DCL and AG).
Work in Zurich was supported by the Swiss National Science
Foundation through the NCCR pool MaNEP.

\end{document}